\documentstyle[epsf,twocolumn]{jpsj}
\catcode`\@=11
\def\runauthor{Masatoshi Imada}
\def\runtitle{Superconductivity Driven by the Interband Coulomb Interaction
and Implications for the Superconducting Mechanism of MgB$_2$}
\def\eqlt{\mathrel{\mathpalette\@vereq<}}  
\def\eqgt{\mathrel{\mathpalette\@vereq>}}  
\def\@vereq#1#2{\lower2.5pt\vbox{\baselineskip0pt \lineskip-.5pt
    \ialign{$\m@th#1\hfil##\hfil$\crcr#2\crcr{=}\crcr}}}
\newcommand{\simle}{\ \raise.3ex\hbox{$<$}\kern-0.8em\lower.7ex\hbox{$\sim$}\ }
\newcommand{\simge}{\ \raise.3ex\hbox{$>$}\kern-0.8em\lower.7ex\hbox{$\sim$}\ }
\begin{document}
\title{Superconductivity Driven by the Interband Coulomb Interaction
and Implications for the Superconducting Mechanism of MgB$_2$}
\author{Masatoshi Imada}
\inst{Institute for Solid State Physics, University of Tokyo,
5-1-5 Kashiwanoha, Kashiwa, Chiba 277-8581, Japan}
\recdate{\today}

\abst{
Superconducting mechanism mediated 
by interband exchange Coulomb repulsion is examined in an extended 
two-band Hubbard models with a wide band crossing the Fermi level 
and coexisting with a narrower band located at
moderately lower energy.  We apply newly developed path-integral renormalization 
group method  to reliably calculate pairing correlations.  
The correlation shows marked enhancement at moderate 
amplitudes of the exchange Coulomb repulsion taken smaller
than the on-site repulsion for the narrower band.  
The pairing symmetry is s-wave while it has unconventional 
phases with the opposite sign between the order 
parameters on the two bands, in agreement with the 
mean-field prediction.  Since the band structure of 
recently discovered superconductor MgB$_2$ shares basic 
similarities with our model, we propose that the present 
results provide a relevant clue for the understanding of the 
superconducting mechanism in MgB$_2$ as well as in this class 
of multi-band materials with 
good metallic conduction in the normal state.}

\kword
{MgB$_2$, pairing mechanism, 
}
\maketitle
Mechanisms of superconductivity have been a long-sort issue since the
discovery in 1911 but are still under active studies with aiming at novel 
mechanisms and searching for materials with higher transition
temperatures.
In particular, mechanisms driven by the electron-electron interaction 
in general are widely and intensively studied issue since the discovery of the
high-Tc cuprate superconductors~\cite{Bednortz}, although there exist not many
superconducting compounds that have been established as driven by the
electron-electron Coulomb interaction. Theoretical studies have 
intensively focussed on the single-band correlated 
electron models.
In this letter we discuss a different possibility of a high-temperature
superconductor by focussing on a wide metallic band coupled 
by a moderate amplitude of exchange Coulomb interaction to the narrower coexisting band or localized orbitals.
 
When two different bands or orbitals are located near the Fermi level in metals,
several completely different physics may appear.  One is the Kondo effect 
and the heavy fermion behavior where the hybridization and
antiferromagnetic spin exchange interaction between localized and itinerant 
bands play a role for the spin singlet formation and heavily renormalized 
effective mass of the quasiparticles in lattice.  On the other hand, if the 
hybridization is vanishing or small between the two orbitals, while the 
exchange Coulomb interaction becomes relatively important, the Hund's rule 
works efficiently and the atomic high-spin state generically appears, if 
the level difference between the two atomic orbitals, $\Gamma$ is 
relatively small.  With this background, the double exchange mechanism 
works in realizing the ferromagnetic ground state in metals. The high-spin 
state appears especially when the ratio between the onsite Coulomb 
interaction for the lower-level orbital, $U$, and $\Gamma$, namely 
$U/\Gamma$ is large.  If the interorbital on-site repulsion $V$ is 
taken into account, this rough estimate of the ratio is replaced 
with $(U-V)/\Gamma$.

However, if $(U-V)/\Gamma$ becomes relatively small or comparable, the 
low-spin state may be stabilized and the Hund's rule coupling becomes 
irrelevant.  Near this stabilization point, $(U-V)/\Gamma\le 1$,   another 
term of the exchange Coulomb interaction between the lower and higher 
orbitals may work for the stabilized low-spin ground state. This term of 
the exchange transfers the up and down spin electrons on the same 
orbital(band) to the other one in pair.  In contrast to the 
ferromagnetism driven by the Hund's rule term, 
we show that this pair transfer part of the exchange interaction 
strongly favors the singlet superconducting order in an appropriate 
range of the parameter space.

In 1963, using the BCS approximation, Kondo~\cite{Kondo} considered the
pairing mechanism assisted by the interband exchange Coulomb interaction 
between $s$ and $d$ bands for the mechanism of the superconductivity
of transition metals.   Yamaji~\cite{Yamaji1,Yamaji2} also considered the similar mechanism 
in a specific condition of two-band Hubbard models. 
A similar interband mechanism was also examined in strongly correlated 
models by the authors~\cite{ImadaKohno}. 

Here, we reexamine the 
mechanism driven by the exchange Coulomb interaction and show by a newly 
developed numerical technique that this mechanism works for simple 
metal compounds mixtured with elements with a localized orbital (or 
relatively narrow band) located moderately below the Fermi level.  Its implication for the 
mechanism of the superconducting transition in MgB$_2$ recently discovered by Akimitsu et al.~\cite{Akimitsu}  above
temperature 39K is an important subject of discussions in this letter.
We discuss that this mechanism may drive or substantially enhance 
the superconducting transition of 
MgB$_2$.
 
Our Hamiltonian
\begin{eqnarray}
{\cal H} & = & {\cal H}_1+{\cal H}_2+{\cal H}_3+{\cal H}_4 \label{eq:1}
\end{eqnarray}
consists of the conduction band ${\cal H}_1$ with the wide and narrow bands or localized 
level part with the on-site Coulomb repulsion terms ${\cal H}_2$ and 
${\cal H}_3$ and with their interorbital Coulomb interaction
${\cal H}_{4}={\cal H}_{4E}+{\cal H}_{4P}+{\cal H}_{4V}$:
\begin{eqnarray}
{\cal H}_1 & = & -\sum_{\langle
ij\rangle,l,\sigma}t_{lij}(c^{\dagger}_{li\sigma}c_{lj\sigma}+{\rm H.c}) \nonumber\\
&&\ \ \ \ \ \ \ -\mu N_1-(\mu +\Gamma)N_2 \label{eq:2a} 
\end{eqnarray}
\begin{eqnarray}
{\cal H}_2 & = & U_1\sum_i n_{1i\uparrow}n_{1i\downarrow} 
\label{eq:2b} 
\end{eqnarray}
\begin{eqnarray}
{\cal H}_3 & = & U_2\sum_j n_{2j\uparrow}n_{2j\downarrow} 
\label{eq:2b2} 
\end{eqnarray}
\begin{eqnarray}
{\cal H}_{4E} & =& -J\sum_{{\scriptstyle i}{\sigma\sigma'}}c^{\dagger}_{1i\sigma}c_{1i\sigma'}
c^{\dagger}_{2i\sigma'}c_{2i\sigma} \label{eq:2c}
\end{eqnarray}
\begin{eqnarray}
{\cal H}_{4P} & =& J\sum_{{\scriptstyle i}{\sigma\sigma'}}
(1-\delta_{\sigma\sigma'})(c^{\dagger}_{1i\sigma'}
c^{\dagger}_{1i\sigma}c_{2i\sigma}c_{2i\sigma'}+ {\rm H.c.}) \label{eq:2d}
\end{eqnarray}
\begin{eqnarray}
{\cal H}_{4V} & = & V\sum_j n_{1j}n_{2j} 
\label{eq:2e} 
\end{eqnarray}
\begin{eqnarray}
N & \equiv & N_1+N_2,\label{eq:3a}
\end{eqnarray}
\begin{eqnarray}
N_1 & \equiv & \sum_{i\sigma} n_{1i\sigma},\label{eq:3b}
\end{eqnarray}
\begin{eqnarray}
N_2 & \equiv & \sum_{l\sigma} n_{2l\sigma},\label{eq:3c}
\end{eqnarray}
In Eq.(\ref{eq:1}-\ref{eq:3c}), the annihilation (creation) operator of the
orbital $l (l=1,2)$ at the site $i$ with the spin $\sigma$ are represented
by $c_{li\sigma} (c^\dagger_{li\sigma})$.
In this letter, we assume that $U_1$ is small by considering a wide
metallic band 1 while $t_{2ij}$ is smaller than $t_{1ij}$.

For the region characterized roughly by $U_2-V\le \Gamma -J$, the atomic ground 
state of the orbital 2 is fully filled singlet.  However, when $J$ becomes not too
smaller than $\Gamma$, the singlet pair can transfer to the conduction band 
1 through the term (\ref{eq:2d}) and it generates a charge fluctuation with 
a singlet cloud around a 
site of the local orbital 2.  In this letter we show that the
overlap of such singlet clouds formed around the neighboring atomic
orbitals 2 may drive the delocalization of the singlet pair and lead to
coherent ground state with the superconducting symmetry breaking.  
Such formation of itinerant singlet pairs survives, as we show later, 
when the orbitals 2 form a band through their mutual transfer.
If we take the region with the low-spin ground state, the terms
(\ref{eq:2c}) and (\ref{eq:2e}) do not play relevant roles and we neglect these terms in
the following discussion for simplicity.

Here we first neglect the dispersion of the lower-level band, $t_2$ in
(\ref{eq:2a})
and the Coulomb repulsion $U_1$ within the conduction band in (\ref{eq:2b}) for
simplicity and discuss their effects later. When we take the mean field approximation with the uniform superconducting
order parameter
\begin{eqnarray}
{\Xi}_1 & = & \frac{J}{N}\sum_{i}\langle
c^{\dagger}_{1i\uparrow}c^{\dagger}_{1i\downarrow}\rangle \label{eq:4a} 
\end{eqnarray}
\begin{eqnarray}
{\Xi}_2 & = & \frac{J}{N}\sum_{i}\langle
c^{\dagger}_{2i\uparrow}c^{\dagger}_{2i\downarrow}\rangle, \label{eq:4b} 
\end{eqnarray}
the selfconsistent equation for the order parameters obtained from the
decoupling of the term (\ref{eq:2d}) for $t_{2ij}=0$ is given by
\begin{eqnarray}
2{\Xi}_1 & = & \Xi_2J/\sqrt{\Xi_2^2+\Gamma^2} \label{eq:5a} 
\end{eqnarray}
\begin{eqnarray}
2{\Xi}_2 & = & -U_2\Xi_2/\sqrt{\Xi_2^2+\Gamma^2}
           \nonumber\\
&&\ \ \ \ \ \ \ + \Xi_1J\int^{W_1/2}_{-W_1/2}{\rm d}\epsilon
         \frac{D_1(\epsilon)}{\sqrt{\epsilon^2+\Xi_1^2}} \label{eq:5b}
\end{eqnarray}
where the density of states of the conduction band 1 is
 $D_1(\epsilon)$ with the dispersion $\epsilon(k)=\sum_{i,j}t_{1i,j}\exp(ik(r_i-r_j))-\mu$ and 
$W_1$ is the bandwidth of the band 1 in proportion to $1/t_1$. 
When the density of states is replaced with the rectangular constant one
for the conduction band for simplicity, namely,  
$D_1(\epsilon) = 1/W_1$, the 
selfconsistent solution at temperature $T=0$ in the weak coupling BCS limit 
is given by
\begin{eqnarray}
\Xi_1 & \sim  & W_1\exp[-\frac{2W_1}{J^2}(U_2+\Gamma/x)]. \label{eq:6}
\end{eqnarray}
Here $x$ denotes the number of the orbital 2 in a unit cell and if the orbital 2 forms a
band through nonzero $t_{2i,j}$, basically $x$ may be replaced with 
the integrated density of states of this narrower band with the width $W_2$.  

An important point of this pairing mechanism is that the prefactor
 in Eq.(\ref{eq:6}) is determined from the bare bandwidth of the wide 
band 1. 
In addition, the argument of the exponential function may have order unity when  we may assume that
$J, U, \Gamma$, and $W_1$ would have comparable values.  
This may necessarily requires the consideration in a strong coupling region. However,
here we took a BCS weak-coupling approach as a starting point and by 
assuming that $W_1$ is the largest parameter among the energy scales.
The strong coupling correction is left for further studies.

Even more important point of this mechanism is that the pairing order parameters on the bands
1 and 2 have the opposite sign, namely 
$\sum_{i,j}\langle c^{\dagger}_{1i\sigma}c^{\dagger}_{1i-\sigma}c_{2j-\sigma}c_{2j\sigma}\rangle<0$
because the pairing is mediated by {\it repulsive} interband exchange interaction.
This anisotropy may enable us to distinguish the present pairing 
mechanism from that of the conventional phonon-mediated BCS theory.

\begin{figure}[tb]
\begin{center}
\epsfxsize=6.2cm
$$\epsffile{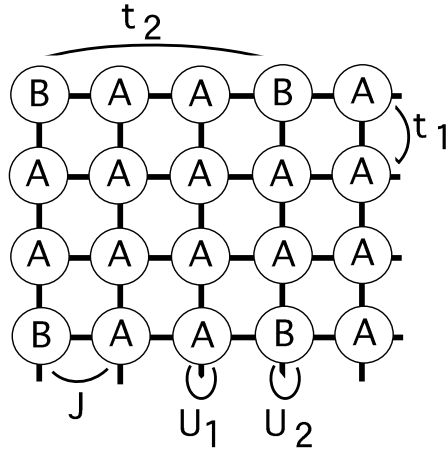}$$
\end{center}
\caption{The lattice structure employed in our numerical simulation.}
\label{Fig.1}
\end{figure}

This weak coupling mean-field picture is complemented by numerical study of the
Hamiltonian (\ref{eq:1}) in the following.  We employ recently 
developed path-integral
renormalization group method\cite{Kashima} to perform a controlled calculation in the
ground state.  Since our basic picture applies to any type of lattice
structure, we first employ a numerically feasible square lattice for
simplicity and introduce orbital 2 by replacing a fraction of the lattice
points (denoted by B in Fig.~\ref{Fig.1}) regularly and assign orbtal 1 to the 
rest of the lattice sites (denoted by A in Fig.~\ref{Fig.1}).  We have introduced 
the B sites periodically in $x$ and $y$ directions with the period $l_B$. 
The wide band 1 is formed from the nearest neighbor transfers $t_1$ between 
the sites A while the narrower band is formed from the nearest neighbor 
transfer $t_2$ between the sites B. 
The unit cell of this lattice becomes $l_B \times l_b$ and it causes
a folding of the Brillouin zone with multi-band structure instead of the band 1.  However it does not alter our basic argument especially near the Fermi level.  The transfer between 
A and B sites is switched off.  
We also introduce an onsite Coulomb repulsion $U_2$ for the sites B 
(on the band 2)
while the onsite Coulomb interaction for the free-electron like 
band 1, $U_1$ is taken small.  The exchange Coulomb interaction 
$J$ given by Eq.(\ref{eq:2d}) is introduced between
nearest neighbor A and B sites. The atomic level difference between 
the sites A and B is $\Gamma$.  

\begin{figure}[tb]
\begin{center}
\epsfxsize=8.8cm
$$\epsffile{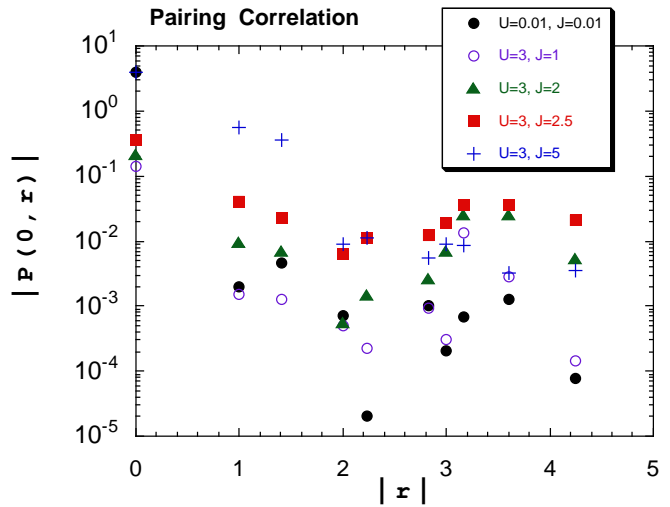}$$
\end{center}
\caption{The absolute amplitude of the pairing correlation $|P|$ as a function of distance for an extended  
two-band Hubbard model.  The choices of $J$ are $J=1,2,2.5$ and $5$.
These results are compared with the practically noninteracting system
with $U_1/t_1=U_2/t_1=J/t_1=0.01$
The pairing correlation shows marked enhancement in a range of the interband 
exchange Coulomb interaction $J$. See the text for details.}
\label{Fig.2}
\end{figure}

We have calculated the pairing correlation 
between the A and B sites as well as other combinations of the pairing 
between two A sites and between two B sites.
Here, the pairing correlation is defined by 
\begin{eqnarray}
P_{lm}(i,j) & = & \langle c^{\dagger}_{li\sigma}c^{\dagger}_{li-\sigma}
c_{mj-\sigma}c_{mj\sigma}\rangle
             \label{eq:7}
\end{eqnarray}
We fix $\Gamma/t_1=U_2/t_1=3, t_2/t_1=0.3$ and $U_1/t_1=0.2$ where the 
largest effect of $J$ is expected in the region of comparable $U_2$ and 
$\Gamma$ as we discussed above.  In a realistic situation, $J$ has to 
be smaller than $U_2$ and we show $J$ dependence of the pairing correlation 
in Fig.~\ref{Fig.2}.  

Figure~\ref{Fig.2} shows the amplitude of the pairing correlation between a site on 
the A sublattice (with the coordinate (0,0)) and all the other sites 
$(i,j)$ including A and B for a 6 by 6 square lattice with the periodic 
boundary condition, $l_B=3$ and the electron density $n=0.5$ per site. 
The distance between (0,0) and $(i,j)$ is given by $r$.
The interband pairing correlations $P_{12}$ has mostly negative values as we expect. 
  The overall correlation
clearly shows marked enhancement for $J$ roughly larger than $U_2/2$.
The enhancement is seen in a comparison with a practically noninteracting case with
$U_1/t_1=U_2/t_1=J_1/t_1=0.01$ with the same band parameters
$t_2/t_1=0.3$ and $\Gamma/t_1=3$ as before.
This implies that the pairing may be driven in a realistic range of the 
parameters when a wide band is coupled by the exchange Coulomb interaction to a narrower band and the 
hybridization between these two bands is vanishing.  

The result also shows that the short-ranged part of the pairing correlation
becomes larger simply with the increase in $J$ while the 
longer-ranged part has a maximum at an intermediate value of $J$ below
$U_2$.  This is a natural consequence because the strongly coupled singlet is tightly 
formed at larger $J$. It bears more local character and an effective
hopping amplitude of the singlet becomes smaller when $J$ becomes too large as 
is seen for an unrealistic choice of $J/t_1=5>U_2/J$.  

Although the size scaling is difficult for the moment,  similar enhancements are also seen in larger lattice sizes.  When the electron
correlations become small, the scale of the enhancements becomes 
small accordingly while the enhancement itself seems to universally exist if the ratios of $U_1, J$ and $\Gamma$ are retained. 
The overall feature of the enhancement is in qualitative agreement
with that expected from the parameter dependence of the mean-field result  Eq.(\ref{eq:6}).  

We next discuss the implication of this result for the mechanism of
superconductivity discovered in MgB$_2$.  The crystal of 
MgB$_2$ has so called AlB$_2$-type hexagonal structure with an alternate stacking of Mg 
and B honeycomb layers.  The band structure~\cite{Harima,Ivanovskii,Kortus} suggests that degenerate 
bands with two-dimensional anisotropy  are formed from B 2$p_{\sigma}$ band
and its bonding part crosses the Fermi level near the top of the band.
In the band structure calculations, the center of the B 2$p_{\sigma}$ bonding band  is 1-2 eV below the Fermi
level and two cylindrical hole Fermi surface appears around the axis through $\Gamma$ to A points due to the bandwidth of the order of 
4 eV. A rather separate free-electron like band also crosses the Fermi
level, which is formed from the hybridization of Mg $sp$ and B $p_z$
bands and has a bandwidth larger than 10 eV.  The Mg $sp$ band is 
expected to have a rather small electron correlation while B $p_{\sigma}$ band may
have a moderate correlation due to dominant $p$ character confined in the
two-dimensional honeycomb layer.  Because of their symmetries, these
two bands do not seem to be appreciably hybridized and are clearly separated near the Fermi level.  This band structure is essentially similar to our model with the band 1 (corresponding to Mg$sp$-B$p_z$) 
and the band 2 (B-$p_{\sigma}$).   Although the 
amplitude of the electron correlation is not well known in this 
compound, the choice of the parameters in our calculations in
comparison with plausible values of parameters for MgB$_2$ does not
seem to be unrealistic and provides a good starting point in the present
stage. 
Our calculated result shows that the pairing enhancement 
becomes optimized for comparable $U$ and $\Gamma$ while it is 
suppressed, for example, for too large $\Gamma$.  This is consistent with the 
fact that AlB$_2$ does not show superconductivity because 
the B $p_{\sigma}$ level is deep in AlB$_2$. On the other hand, if $\Gamma$
becomes too small, the Hund's rule coupling starts working and the ferromagnetic
state becomes more stable.  In a series of the binary compounds of boronides,
we expect that the ferromagnetic and superconducting phase may appear rather 
close each other with competitions.
 
In this compound, phonon modes of Boron oscillation may have relatively
high frequencies and may take a strong electron-phonon coupling which may 
support the phonon-mediated conventional 
pairing~\cite{Kortus,An}.  Our proposal gives an alternative possibility 
or may play a complementary role.  In comparison with the phonon mechanism, our
mechanism contains a specific prediction where the phases of the 
order parameter on the B $p_{\sigma}$ band and B $p_z$ bands hybridized with 
Mg $sp$ should have the opposite sign.  This is a crucial test for the role 
of the interband Coulomb exchange as a
driving force of the superconductivity.   This may be tested in a 
carefully designed junction to pick up the opposite phase.  
Although the pairing 
symmetry is s-wave as supported also experimentally~\cite{Kitaoka} , 
our theory suggests that it is not a simple BCS isotropic pairing.

In summary, we have analyzed a superconducting mechanism mediated 
by the interband Coulomb exchange repulsion.  The numerical simulation
results suggest enhancements of pairing for the combination of wide and 
narrower bands crossing the Fermi level without mutual hybridization but 
with a moderate amplitude of the exchange Coulomb interaction.  We have
proposed that this mechanism may be relevant to the occurrence of 
superconductivity recently discovered in MgB$_2$.

\section*{Acknowledgements}
The author would like to thank J. Akimitsu for fruitful discussions on detailed 
experimental results and presenting the data to the author 
in prior to the publication.  The author
also thanks H. Harima for illuminating discussions and data for the
band structure.  The fruitful conversation with K. Kusakabe is also 
acknowledged.  This
work is supported by a Grant-in-Aid for ``Research for the Future" Program from
the Japan Society for the Promotion of Science under the project
JSPS-RFTF97P01103.



\end{document}